\documentstyle[psfig,aps,floats,eqsecnum,prd,preprint,tighten]{revtex}

\newcommand{\be}{\begin{equation}}
\newcommand{\ee}{\end{equation}}
\newcommand{\bea}{\begin{eqnarray}}
\newcommand{\eea}{\end{eqnarray}}
\newcommand{\bref}[1]{(\ref{#1})}
\newcommand{\ket}[1]{\left|{#1}\right>}
\newcommand{\bra}[1]{\left<{#1}\right|}
\newcommand{\expect}[1]{\left<{#1}\right>}
\newcommand{\hconj}{(\mbox{h.\ c.})}

\newcommand{\Lagf}{{\cal L}_{\text{fermions}}}
\newcommand{\ms}{M_{\text{S}}}
\newcommand{\mB}{M_{\text{B}}}
\newcommand{\mt}{\widetilde M}
\newcommand{\sig}[1]{\sigma^{#1}}
\newcommand{\pd}[1]{\partial_{#1}}

\newcommand{\eps}{\epsilon}
\newcommand{\Phistr}{\Phi_{\text{S}}}
\newcommand{\mL}{m_{\! L}}
\newcommand{\mew}{m_{\text{ew}}}

\newcommand{\bk}{{\bf k}}
\newcommand{\bx}{{\bf x}}
\newcommand{\sstR}{{\scriptscriptstyle 2\bf R}}
\newcommand{\sGUT}{{\scriptscriptstyle \text{GUT}}}
\newcommand{\bR}{{\bf R}}
\newcommand{\bmat}[1]{\left(\begin{array}{{#1}}}
\newcommand{\emat}{\end{array}\right)}

\begin{document}

\title{\bf Cosmic String Current Stability}

\author{Stephen C. Davis\footnote{S.C.Davis@swansea.ac.uk}, 
 Warren B. Perkins\footnote{w.perkins@swansea.ac.uk}}
\address{Department of Physics, University of Wales Swansea,\\
\em  Singleton Park, Swansea, SA2 8PP, UK}
\author{Anne-Christine Davis\footnote{A.C.Davis@damtp.cam.ac.uk}}
\address{Department of Applied Mathematics and Theoretical Physics,\\
\em University of Cambridge, Cambridge, CB3 9EW, UK}

\maketitle

\begin{abstract}

The stability of fermionic charge carriers on cosmic strings is
considered. We show that neutral fermion currents in cosmic strings are
always chiral or time-like, in contrast to the case 
of bosonic currents. The spectrum of bound states on an abelian
SO(10) string is determined both before and after the electroweak
phase transition. We determine the mass acquired by the zero mode at
this transition. A range of charge carrier scattering processes are
considered and corresponding decay rates calculated. Couplings between
the carriers and the electroweak sector generate scattering from the
plasma which can thermalise some currents. If the zero mode is
isolated from the electroweak sector, then it
survives. Current--current scattering is considered, but found to be
unimportant in realistic settings where the string density is low.

\end{abstract}

\setcounter{page}{0}
\thispagestyle{empty}

\vfill

\begin{flushright} DAMTP/99-168 

 SWAT/229 \end{flushright}

\vfill

\newpage

\section{Introduction}

Although much of the evolution of the Universe is well understood,
there are still many cosmological phenomena for which a completely
satisfactory explanation has yet to be found. Topological defects, such
as cosmic strings, could provide mechanisms for structure formation,
CMB anisotropy, and high energy cosmic rays~\cite{rays}. Such defects form
in many realistic unified theories. 

In the past it has been difficult to evaluate the usefulness of such
ideas due to a lack of data. This is now changing, and predictions of
CMB anisotropies from simple cosmic string models have been
made~\cite{Turok}. While these predictions show poor agreement with
the observations, they  do not take into account the full physics of
strings models.  Indeed, recent analysis which includes the effect of
particle production as an energy loss mechanism from the string network
show much improved agreement with data~\cite{joao&nathalie}.
One possibility is that the strings carry conserved
currents~\cite{supercond}. These currents will alter the evolution of
a string network, which could lead to better agreement with
observation. Indeed, an analytic analysis showed that a much denser 
string network results for electromagnetically coupled strings~\cite{kostas}.
However, one significant implication of conserved currents
is that they can stabilise loops of string. If persistent,
these stabilised loops or `vortons'~\cite{vorton,vorton2}, can easily
dominate the energy density of the Universe, placing stringent
constraints on the parameters of the model. An analysis has been made of
the implications of this for particle physics models predicting current
carrying strings~\cite{brandon&acd}.

Fermions are a natural choice for the charge carriers of such
currents. Fermion zero modes exist in a wide class of cosmic string
models. The fermions can be excited and move along the string, thus
resulting in a current. Consequently, fermion conductivity occurs
naturally in many supersymmetric and grand unified theories, such as
SO(10). Most attention has been given to massless, chiral currents, as
these are most  likely to be stable. In this paper space- and
time-like currents are also considered. These naturally occur in the
bosonic models of superconducting strings~\cite{boscurr}. In
section~\ref{pos} we show that space-like fermion bound states do not
exist in the string core in any cosmic string model. We contrast this
with the situation for bose current carriers. In
section~\ref{spectrum} the spectrum of time-like currents is
investigated for an abelian string model. We determine the complete
spectrum of fermion modes, including both zero modes and massive bound
states in our analysis. Whilst traditionally, zero modes are
considered for current carriers, low-lying bound states can also carry
currents on strings. This analysis is applied to the SO(10) model. The
effect of the electroweak phase transition on fermion currents is
investigated in section~\ref{EWtran}.

One criticism of fermion currents is that, unlike scalar boson
currents, they are not topologically stable. It is possible that they
could decay, either directly into particles off the string, or through
interactions with the surrounding plasma. If the decay rate is too
high, currents will not last long enough to have any significant
effect. On the other hand if the decay rate is too low the Universe
could become vorton dominated. Such decays are examined in
section~\ref{decay}. We first examine the effect of plasma scattering on
the string current. We show that this process can remove current carriers
close to the phase transition, but not otherwise. We also consider the
effect of current--current scattering and apply our analysis to a network
of strings shortly after the phase transition. We 
show that this also has a negligible effect on the current stability.
The results are summarised in section~\ref{sum}.

\section{Positivity Condition}
\label{pos}

\newcommand{\dslash}{\,/\!\!\!\!D}

In this section we consider the possible fermionic currents that can
be carried by a string. Consider the following fermionic Lagrangian
for a set of fermions $\psi_i$:
\be
{\cal L}=\overline\psi_i i\dslash \psi_i
+\bigl[\overline\psi_i\phi_jm^D_{ijk}\psi_k +\hconj\bigr]
+\bigl[\overline\psi_i\phi_jm^M_{ijk}\psi^c_k +\hconj\bigr] \ .
\label{lag1}
\ee
Here, $\psi^c_i= C \overline\psi^T_i$, and $C$ is
the charge conjugation matrix.

For shorthand we write, $M^D_{ik}=\phi_j m^D_{ijk}$ and 
$M^M_{ik}=\phi_j m^M_{ijk}$. The Dirac equation is then,
\be
\gamma^0i\dslash \psi_i
+\bigl[\gamma^0 M^D_{ik} + M^{D\dagger}_{ki} \gamma^0 \bigr]\psi_k
+\bigl[\gamma^0 M^M_{ik}
       - \gamma^{0T} C^T M^{MT}_{ki} C^{-1} \bigr]\psi^c_k=0 \ .
\ee
For concreteness we work  with the Dirac representation of the gamma
matrices. In this basis $\gamma^0$, $\gamma^3$ and  $C=i\gamma^2 \gamma^0$
are real. Combining the Dirac equation and its charge
conjugate into a single matrix equation we find,
\be
\left[\hat H_0 + \gamma^0 \gamma^3 \hat P_3 + \hat H_m\right] 
\pmatrix{\psi_i \cr \psi^c_i \cr} = 0 \ ,
\ee
where
\begin{displaymath}
\hat H_0=\pmatrix{ iD_0 & 0 \cr
		      0 & iD^*_0 \cr} \ 
,\qquad
\hat P_3=\pmatrix{ iD_3 & 0 \cr
                      0 & iD^*_3 \cr} \ ,
\end{displaymath}
\be
\hat H_m=\pmatrix{
i\gamma^0 \gamma^a D_a + \gamma^0 M^D_{ik} + M^{D\dagger}_{ki}\gamma^0 &
\gamma^0 M^M_{ik} + \gamma^0 C^\dagger M^{MT}_{ki} C \cr
M^{M\dagger}_{ki} \gamma^0 + C^\dagger M^{M*}_{ik} C \gamma^0 &
i\gamma^0 \gamma^a D^*_a 
+ \gamma^0 C^\dagger M^{DT}_{ik} C + C^\dagger M^{D*}_{ki} C \gamma^0 \cr} \ .
\ee
The index $a$ runs over the values 1 and 2.

The operators $\hat H_0$ and $\hat P_3$ are clearly hermitian. If we
restrict the mass matrices so that the Lagrangian contains only
Lorentz scalar or pseudoscalar terms then  $\hat H_m$ is also
hermitian, and $\{\gamma^0\gamma^3,\hat H_m\}=0$.

Let a state $\ket{\psi}$ now represent the vector
$(\psi_i, \psi_i^c)^T$. 
For a simultaneous eigenstate of $\hat H_0$ and $\hat P_3$ we have,
\be
\bra{\psi}\hat H_0 \hat H_0\ket{\psi}
=\bra{\psi}(\gamma^0\gamma^3\hat P_3
+\hat H_m)(\gamma^0\gamma^3\hat P_3 +\hat H_m)\ket{\psi} \ , 
\ee
or
\be
w^2=k_3^2 +\bra{\psi}\{\gamma^0\gamma^3\hat P_3,\hat H_m\}\ket{\psi} 
+\bra{\psi}\hat H_m\hat H_m \ket{\psi} \ .
\ee
Now
\be
\{\gamma^0\gamma^3\hat P_3,\hat H_m\}=
[\hat P_3,\gamma^0\gamma^3\hat H_m]
+\{\gamma^0\gamma^3,\hat H_m\}\hat P_3 \ .
\ee
For the mass matrices we are considering, the second term vanishes.
Since $\ket{\psi}$ is an eigenstate of $\hat P_3$,
\be
\bra{\psi}[\hat P_3,\gamma^0\gamma^3\hat H_m]\ket{\psi}
=\bra{\psi}[k_3,\gamma^0\gamma^3\hat H_m]\ket{\psi}=0 \ .
\ee
Thus the anticommutator term vanishes and we have
\be
w^2=k_3^2 +\bra{\psi}\hat H_m\hat H_m \ket{\psi} \ .
\ee

As we have seen, $\hat H_m$ is hermitian, thus the expectation value
is a weighted sum of the squares of real eigenvalues, i.e.\ it is
positive definite:
\be
w^2 \ge k_3^2 \ .
\ee 

Thus for scalar and pseudoscalar mass terms, we see that there are no
space-like fermions on the string and the total energy-momentum of the
charge carriers is null or time-like.  We can interpret this result
physically as guaranteeing the stability of the string against the
spontaneous formation of a fermionic condensate.  This contrasts
sharply with the bosonic case~\cite{boscurr}, where the   bare string
is unstable to the spontaneous formation of a bosonic condensate.
This not only  allows the formation of a uniform, static condensate,
but also space-like excitations, i.e.\ static condensates whose phases
wind along the string.  The presence of such states relies on two
features of the bosonic model which are not present in the fermionic
case;  the non-linearity of the equations of motion  and the effective
scalar mass squared being negative in the string core.

Of course if there are oppositely charged fermion  carriers moving in
different directions along the string, the charge  current can be
space-like, but the energy-momentum is still null or time-like.

\section{Bound States in the Abelian String Model}
\label{spectrum}

Having shown that only light- and time-like fermions exist on cosmic
strings, we will now determine the complete spectrum of currents in a
simple U(1) string model. A model with a suitable symmetry breaking
has the Lagrangian
\be
{\cal L} = (D_\mu \Phistr)^\dagger D^\mu \Phistr 
- \frac{1}{4}F_{\mu \nu}F^{\mu \nu} 
- \frac{\lambda}{4}\left(|\Phistr|^2-\eta^2\right)^2
\label{Lagboson}
\ee
where $D_\mu \Phistr = (\partial_\mu - ie A_\mu) \Phistr$ and
$F_{\mu \nu} = \partial_\mu A_\nu - \partial_\nu A_\mu$. This is just
the abelian Higgs model.
In the usual vacuum
solution the mass of $\Phistr$ is $\ms=\sqrt{\lambda}\eta$, and the
gauge field $A_\mu$ has mass $M_{\text{V}}= \sqrt{2}e\eta$. The vacuum
is topologically non-trivial, allowing cosmic string
solutions to exist. To simplify the analysis, we consider strings with 
winding number 1. In this case 
$\expect{A_\mu} = \delta^\theta_\mu a(\rho)/er$ and
$\expect{\Phistr}=\eta f(\rho)e^{i\theta}$, where $\rho = \ms r$. The
boundary conditions of the two functions are $f(0)=a(0)=0$ and
$f(\infty)=a(\infty)=1$. The radius of the string is of order $\ms^{-1}$.
 
The model can be extended by adding a Weyl fermion field, $\Psi$. The extra
fermionic part of the Lagrangian is then
\be
\Lagf = \Psi^\dagger i\sig{\mu} D_\mu \Psi 
- \frac{1}{2} ig_\psi \Psi^\dagger \Phistr \Psi^c
+ \frac{1}{2} ig_\psi^* \Psi^{c\dagger} \Phistr^* \Psi \ , 
\label{Lagferm}
\ee
where $\sig{\mu} = (I,\sig{i})$, $\Psi^c = i\sig{2} \Psi^*$ and
$D_\mu \Psi = (\partial_\mu - \frac{1}{2}ie A_\mu) \Psi$. 

The variation of $\Phistr$ means that the cosmic string acts as a
potential well for the fermions. As well as the usual continuum states, 
there will be additional fermion states which exist only on the
string. These are the fermion zero modes and massive bound states. The
fermion field $\hat \Psi$ can be expressed in terms of momentum
eigenstates. The bound states have only $z$- and angular momentum
($k_z$, $n$) since they are restricted to the string core. Apart from the
massive bound states there are also massless chiral solutions. These
have $n=0$ and, as a consequence of their chirality, only travel in
one direction along the string.
\bea
\hat \Psi &=& \sum_{k_z n i} \left( \hat c_{i\bk} U_{i\bk} e^{-iw t}
+ \hat c_{i\bk}^\dagger \tilde{U}_{i\bk} e^{iw t} \right) 
\nonumber \\ &&
{}+ \sum_{k_z>0} \left( \hat c_{0\bk} U_{0\bk} e^{-iw t} 
+ \hat c_{0\bk}^\dagger U^*_{0\bk} e^{iw t} \right) 
+ \sum_\bk (\mbox{continuum states}) \ .
\eea

The index $i$ runs over the different masses of the bound states. It
can be seen that the massless chiral states ($U_{0\bk}$) are real. The
field equations for the spinors $U$ and $\tilde{U}$ can be found by
varying \bref{Lagferm} with respect to $\Psi^\dagger$. The wavefunctions 
are then found to have the form,
\bea
U_{i\bk} &=& \frac{\mt}{\sqrt{2w\pi L}} 
\bmat{c} \sqrt{w + k_z} \chi_1(\rho) \\ 
	i \sqrt{w - k_z} \chi_2(\rho) e^{i\theta} \emat 
	e^{i(k_z z + n\theta)} \nonumber \\
\tilde{U}_{i\bk} &=& \frac{\mt}{\sqrt{2w\pi L}}
	\bmat{c} \sqrt{w + k_z} \chi_3 (\rho) \\
	-i \sqrt{w - k_z} \chi_4 (\rho) e^{i\theta} \emat
	e^{-i(k_z z + n\theta)} \label{chiansM} \\
U_{0\bk} &\propto& \frac{\mt}{\sqrt{\pi L}} \bmat{c} 1 \\ 0 \emat 
e^{-\int \frac{M}{\ms} f + \frac{a}{2\rho} d\rho} e^{ik_z z} \ ,
\label{chians0}
\eea
where $L$ is the length of the string, $\mt^{-1}$ is  the effective
radius of the wavefunctions and $M = |g_\psi| \eta \sim \ms$ is the
fermion mass away from the string.  The spinors are normalised so that
$\int d^3x \, (|U_{i\bk}|^2 + |\tilde{U}_{i\bk}|^2) = 1$ and
$\int d^3x \, |U_{0\bk}|^2 = 1/2$.

The allowed masses of the bound states ($\mB$) can be determined by
finding normalisable solutions for the equations satisfied by the
functions $\chi_i$. In a massive bound state's rest frame these are
\bea
\ms\left(\pd{\rho} -\frac{n}{\rho} + \frac{a}{2\rho} \right) \chi_1
+ \mB \chi_2 + M f \chi_3 = 0&& \label{chi1} \\
\ms\left(\pd{\rho} +\frac{n+1}{\rho} - \frac{a}{2\rho} \right) \chi_2
- \mB \chi_1 - M f \chi_4 = 0&& \label{chi2}\\
\ms\left(\pd{\rho} +\frac{n}{\rho} + \frac{a}{2\rho} \right) \chi_3
+ \mB \chi_4 + M f \chi_1 = 0&& \label{chi3}\\
\ms\left(\pd{\rho} -\frac{n-1}{\rho} - \frac{a}{2\rho} \right) \chi_4
- \mB \chi_3 - M f \chi_2 = 0 &\! .& \label{chi4}
\eea
Of course $n$ must be an integer for the solution to be single valued
and hence physical. Examining the above equations we see that given
one solution, another can be obtained by putting $n \rightarrow -n$,
$\chi_1 \leftrightarrow \chi_3$ and $\chi_2 \leftrightarrow \chi_4$.
Thus the solutions will occur in pairs.

We will use a variation of the shooting method to determine the
allowed values of $\mB$. At large $\rho$ the solutions of
\bref{chi1}--\bref{chi4} have exponential behaviour. Two of them decay
and so are acceptable. In the case of the small $\rho$ solutions, only
two of them give a normalisable state.


\begin{figure}
\begin{picture}(415,300)
\put(5,8){\psfig{file=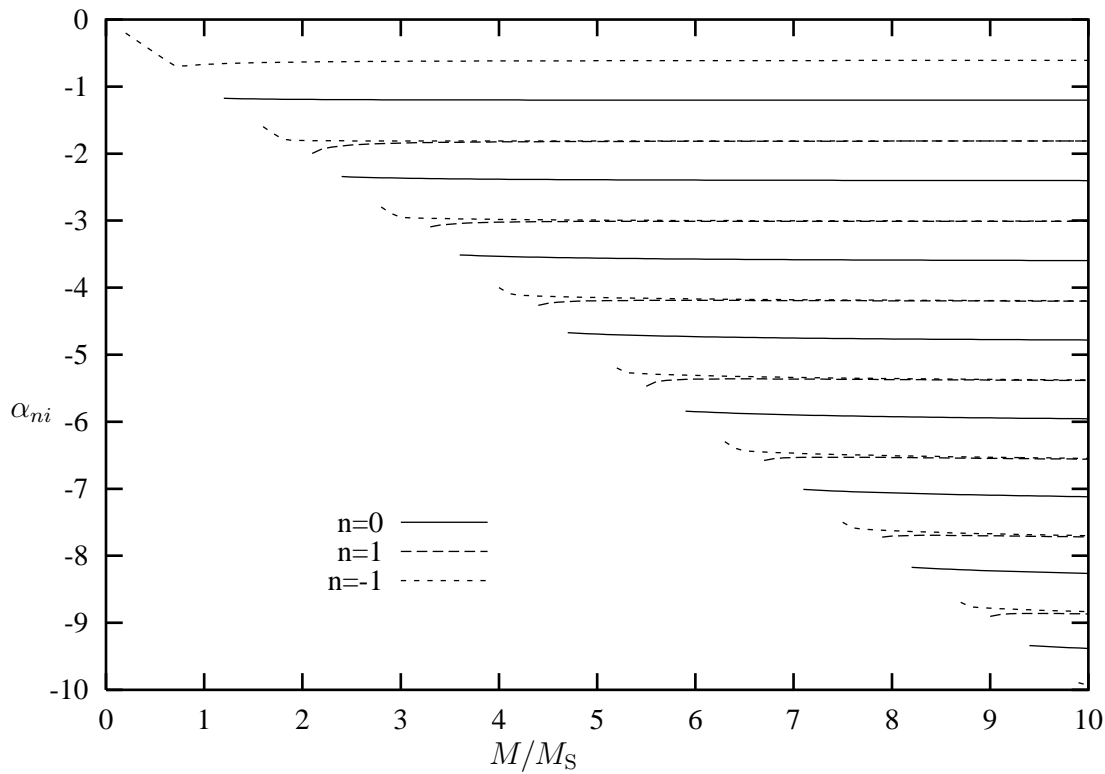}}
\put(10,135){\makebox(0,0){$\alpha_{ni}$}}
\put(200,5){\makebox(0,0){$M/\ms$}}
\end{picture}
\caption{Spectrum of $n=0,\pm 1$ massive fermion bound states.} 
\label{bsspec1 fig}
\end{figure}

Each of these 4 solutions can be numerically extended to some
intermediate value of $\rho$ (of order 1). It can then be seen if any
non-trivial combinations of the large and small $\rho$ solutions match
up there. We find that the values of $\mB$ for which normalisable
solutions exist satisfy
\be
M^2 - \mB^2 = (M + \alpha_{ni}\ms)^2 > 0
\ee
where the $\alpha_{ni}$ are functions of $M/\ms$. Figure~\ref{bsspec1
fig} shows the variation of $\alpha_{ni}$ and the number of bound
states with respect to the fermion to Higgs mass ratio ($M/\ms$). Each
line corresponds to two bound states. For simplicity we have taken the
Higgs and gauge field masses ($\ms$ and $M_{\text{V}}$) to be equal. With
this choice of parameters all solutions have either $\chi_2$ or
$\chi_4$ identically zero. Equations \bref{chi1}--\bref{chi4} can then
be reduced to a second order problem. The solutions with $\chi_4=0$
are shown in the figures. It can be seen from figure~\ref{bsspec1 fig}
that the number of bound states increases with the size of the
off-string fermion mass. For small values of $M$ there are just two
massive bound states in addition to the chiral zero mode.

We find that in the region of parameter space in which $\alpha_{ni}$
is roughly constant, $\mt^2 \approx \ms\sqrt{M^2 - \mB^2} / 2$. For
the massless states, when $\ms=M_{\text{V}}$, this expression for $\mt$ is
exact and can be proved analytically. In this case the two sides of
\bref{chians0} are actually equal.

\begin{figure}
\begin{picture}(435,180)
\put(0,0){\psfig{file=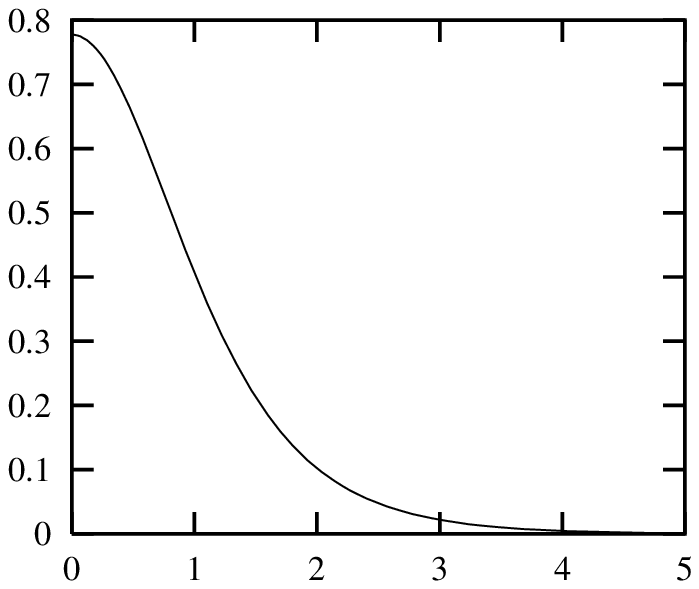}}
\put(210,0){\psfig{file=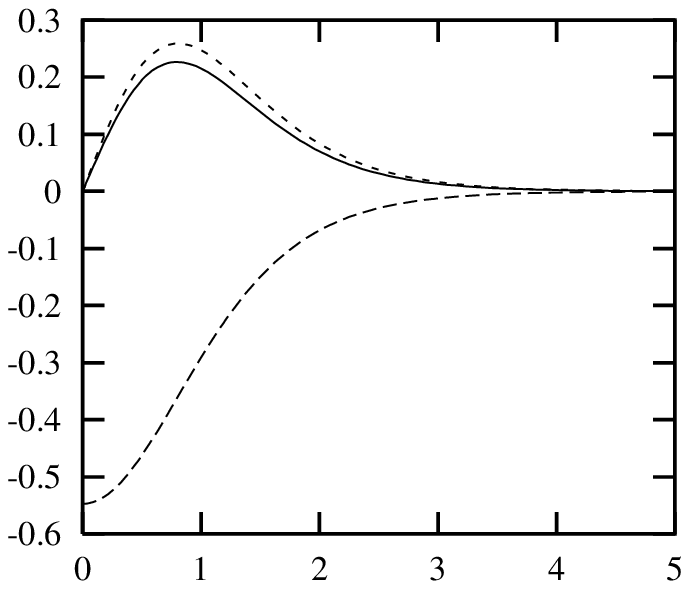}}
\put(130,100){\shortstack[l]{$n=0$ \\ $\mB=0$}}
\put(320,40){\shortstack[l]{$n=-1$ \\ $\alpha = -0.634$
\\ $\mt = 0.826\ms$ \\ $\mB = 1.46\ms$}}
\put(190,10){\makebox(0,0){$\mt r$}}
\put(400,10){\makebox(0,0){$\mt r$}}
\put(290,140){\makebox(0,0){$\chi_1$}}
\put(290,70){\makebox(0,0){$\chi_2$}}
\put(305,155){\makebox(0,0){$\chi_3$}}
\end{picture}
\caption{Zero mode and lowest massive bound state solutions in the abelian
string model with $M/\ms = 2$.}
\label{bssol fig}
\end{figure}

Plots of the solutions for $M/\ms = 2$ are shown in
figure~\ref{bssol fig}. As expected they decay outside the string.
We have thus found the full spectrum of fermion bound states for this model. 

\section{Currents after the Electroweak Phase Transition}
\label{EWtran}

The toy model of the previous section can be embedded in a
phenomologically credible grand unified theory, such as SO(10). Two
suitable symmetry breakings which can give rise to cosmic strings are
\be
\text{SO(10)} \stackrel{\scriptstyle \Phistr}{\longrightarrow}
\text{SU(5)} \times Z_2 \longrightarrow 
  {\cal G}_{SM} \times Z_2 
\stackrel{\scriptstyle \Phi}{\longrightarrow}
\text{SU(3)}_c \times {\rm U(1)}_Q \times Z_2 \ ,
\ee
\be
\text{SO(10)} \rightarrow \text{SU(5)} \times \text{U(1)} 
\rightarrow {\cal G}_{SM} \times \text{U(1)}
\stackrel{\scriptstyle \Phistr}{\longrightarrow} {\cal G}_{SM} \times Z_2 
\stackrel{\scriptstyle \Phi}{\longrightarrow}
\text{SU(3)}_c \times \text{U(1)}_Q \times Z_2
\ee
with ${\cal G}_{SM} = \text{SU(3)}_c \times \text{SU(2)}_L \times 
\text{U(1)}_Y$.
The discrete $Z_2$ symmetry allows the formation of topologically
stable cosmic strings. In this case the string gauge field is a neutral
GUT boson, and $\Phistr$ transforms under the {\bf 126} representation of
SO(10). The electroweak Higgs field $\Phi$ transforms under
the {\bf 10} representation.

The fermion sector of the SO(10) GUT contains all the Standard Model
fermions, and an extra right-handed neutrino ($\nu^c$) for each family. Only
right-handed neutrinos couple to $\expect{\Phistr}$, while neutrinos of
either helicity couple to $\expect{\Phi}$.  

The abelian string's gauge field has a non-trivial effect on the electroweak
Higgs field. The components of $\Phi$ have charges $\pm 1/5$ with
respect to the generator of the GUT string gauge field, so $\expect{\Phi}$
will not be constant in the presence of a string. A non-zero $Z$ field is also
required to give a finite energy solution~\cite{so10EWstr}.

The $\tau$-neutrino mass terms in a string background are
\be
\bmat{cc} \nu^{c\dagger} & \nu^\dagger \emat 
\bmat{cc} M f(\rho) e^{i\theta} & \mew h(\rho) \\ \mew h(\rho) & 0 \emat 
\bmat{c} \nu^c \\ \nu \emat^c \ .
\ee
$\mew \sim 10^2$GeV is the electroweak energy scale and
$M \sim 10^{16}$GeV is the GUT energy scale. For simplicity we will neglect
mixing between different families.  The function $h$ gives the radial
dependence of the component of $\expect{\Phi}$ which couples to the
neutrinos.  The electroweak gauge field has
$\expect{Z_\theta} = -b(\rho)/(\sqrt{40}er)$. The boundary conditions
of $h$ and $b$ are $b(0)=0$ and $h(\infty)=b(\infty)=1$. Inside the
string $h \approx (\mew/M) c=$constant, with $c \sim10^4$~\cite{SCDthesis}.
While the GUT fields take their vacuum expectation values outside the
string (whose radius is of order $M^{-1}$), the electroweak fields
vary over a far larger region with radius of order $\mew^{-1}$.
 
Since $\eps = \mew / M \ll 1$ the neutrino mass eigenvalues outside the
string (or in a vacuum) are approximately $M$ and $\mL=\mew^2/M$. The
corresponding mass eigenstates are then approximately $\nu^c + \eps \nu$ and
$\nu - \eps \nu^c$. This illustrates the
seesaw mechanism~\cite{seesaw}. Although the $\tau$-neutrinos have the same
couplings to the electroweak Higgs field as the top quark, the GUT
Higgs ensures that $\nu^c$ is superheavy and $\nu$ is very light, as
is required to agree with observation. Recent measurements have suggested
that $\nu$ does indeed have a small mass~\cite{numass}.

Once $\Phi$ gains a non-zero expectation value, the $\nu^c$ currents
are no longer solutions of the field equations. We will start by
considering the fate of the massless $\nu^c$ currents. It seems likely
that they mix with a $\nu$ state to give a low mass bound state. We
denote the resulting mass of the bound state by $\mu M$, where the
dimensionless parameter $\mu$ is expected to be very small. Using the
ansatz
\be
\nu^c = \bmat{c} \chi_1(\rho) \cos (\mu M t) \\ 
\chi_2(\rho) e^{i\theta} \sin (\mu M t) \emat 
\ , \ \ \ \nu = \bmat{c} \zeta_1(\rho) e^{-i\theta} \cos (\mu M t) \\
\zeta_2(\rho) \sin (\mu M t) \emat 
\ee
we can look for solutions which reduce to the zero mode solutions of
\bref{chi1}--\bref{chi4} when $\expect{\Phi}=0$. Putting $\ms=M$ for
simplicity, the field equations reduce to
\bea
\left(\pd{\rho} + \frac{a}{2\rho} + f\right) \chi_1 
+ \mu \chi_2  + \eps h \zeta_1 = 0 && \label{ewchi1} \\
\left(\pd{\rho} + \frac{2-a}{2\rho} - f\right) \chi_2 
- \mu \chi_1  - \eps h \zeta_2 = 0 && \\
\left(\pd{\rho} + \frac{10-2b-3a}{10\rho}\right) \zeta_1 
+  \mu \zeta_2 + \eps h \chi_1 = 0 && \\
\left(\pd{\rho} + \frac{2b+3a}{10\rho}\right) \zeta_2 
- \mu \zeta_1 - \eps h \chi_2 = 0 &\! .& \label{ewchi4}
\eea

As with the abelian string, the existence of bound states can be
investigated by examining the large and small $\rho$ solutions of
\bref{ewchi1}--\bref{ewchi4}, and then trying to match them at
intermediate $\rho$. There are two well behaved solutions at small $\rho$
and two at large $\rho$ if $\mu < \eps^2$.

We will attempt to do this by finding approximate solutions in a
simple `top hat' approximation of the string background. This has
$f(\rho)=a(\rho)=\Theta(\rho)$, $b(\rho)=\Theta(\eps\rho)$ and
$h(\rho)=\eps c + (1-\eps c)\Theta(\eps\rho)$, where $\Theta$ is the
Heaviside step function. It is then possible to find analytic
solutions for $\rho > \eps^{-1}$. For smaller $\rho$ the solutions can
be expanded in terms of the tiny parameter $\eps$.

Requiring the approximate solutions to match up at $\rho=1$ and
$\rho=\eps^{-1}$ will give an expression for $\mu$. It is satisfied by
$\mu=\sqrt{2} \gamma c \eps^{16/5}$ with
$\gamma = e\Gamma(6/5,1) - 1/2 \approx 1.6$. $\Gamma$ is the
incomplete gamma function. To leading order in $\eps$, the
corresponding approximate solution is
\bea 
\chi_1 &=& 1 \nonumber \\ 
\chi_2 &=& \frac{c}{\sqrt 2}(\gamma-1) \eps^{16/5} \rho \nonumber \\
\zeta_1 &=& -\frac{c}{2} \eps^2 \rho \nonumber \\
\zeta_2 &=& -\sqrt{2}\eps^{6/5}
\eea
for $\rho < 1$,
\bea
\chi_1 &=& \frac{e^{1-\rho}}{\sqrt \rho} 
+ \eps^4 c\gamma \left( c\frac{5}{4} 
{}_1 \!F_1\!\left(\frac{4}{5},\frac{9}{5};\rho\right)e^{-\rho}\rho^{3/10} 
- 2\eps^{6/5}\frac{e^{\rho-1/\eps}}{\sqrt \rho}\right) \nonumber \\
\chi_2 &=& \sqrt{2} c \left(\Gamma\!\left(\frac{6}{5},\rho\right)
\frac{e^\rho}{\sqrt \rho} -
\frac{\gamma}{2} \frac{e^{1-\rho}}{\sqrt \rho} \right) \eps^{16/5}
+ \sqrt{2} \eps^2 \frac{e^{\rho-1/\eps}}{\sqrt \rho} \nonumber \\
\zeta_1 &=& \eps^2 \left(e\Gamma\!\left(\frac{6}{5},\rho\right) 
- \gamma\right) \rho^{-7/10}\nonumber \\
\zeta_2 &=& -\sqrt{2}\eps^{6/5} \rho^{-3/10}
\eea
for $1 < \rho < \eps^{-1}$, and 
\be
\bmat{c} \zeta_1 - \eps \chi_1 \\ \zeta_2 - \eps \chi_2 \emat
= -\bmat{c} \gamma c \eps^{6/5} \\ \sqrt{2} \emat \eps^{6/5}
\frac{e^{-\eps^2 \rho}}{\sqrt \rho} \ , \ \ \
\bmat{c} \chi_1 + \eps \zeta_1 \\ \chi_2 + \eps \zeta_2 \emat
= O(\eps^{-1/5}) \frac{e^{-\rho+1/\eps}}{\sqrt \rho}
\ee
for $\rho > \eps^{-1}$. ${}_1\!F_1$ is a confluent hypergeometric
function.
 
The mass of the new bound state is $M\mu \sim 10^{-16}$eV. Unlike the
original zero mode, it is a mixture of left- and right-handed
neutrinos. Similarly, the massless $\nu^c_\mu$ and $\nu^c_\tau$
currents will also gain tiny masses at the electroweak phase
transition. Since these states now couple to light off-string states,
they can escape from curved strings~\cite{vorton}. Applying the
arguments in ref.~\cite{vorton} the rate of decay of states with energy
of order $\ms$ is calculated to give a lifetime of order
$10^{-40}$s. This would prevent the Universe becoming vorton
dominated.  In addition the electroweak bound states are spread over a
far greater region than the original $\nu^c$ current, so interactions
with other fields will also be increased. If any massive bound states
survive the phase transition they will also rapidly decay into
left-handed neutrinos, although it seems more likely that they will
simply be changed into free states by the transition.

The above mechanism will work in other theories where zero modes mix
with free massless particles. If there is mixing between two massive
species of right and left moving currents, zero modes can mix to give
a low mass bound state. However, as there is no low mass off-string
state these cannot escape.  Such theories always feature zero modes
moving in opposite directions. These can interact with each other,
which reduces the current. It is still possible for a net current
to persist, and so fermion currents may provide constraints on these
models after all.

If a zero mode is isolated from the electroweak sector, then it
persists. Consequently the resulting currents also persist. In
particular, zero modes in grand unified theories or supersymmetric
theories where there is no coupling between the zero mode and the
electroweak sector, result in currents which are not
destroyed. Generally, such currents give rise to stable vortons,
which constrain the underlying particle physics
theory~\cite{brandon&acd}. This is likely to be particularly true in
D-type supersymmetric theories.

\section{Decay Rates of Charge Carriers}

\label{decay}

\begin{figure}
\begin{picture}(410,100)
\put(10,20){\begin{picture}(120,50)
\put(40,30){\line(-2,1){25}}
\put(40,30){\line(-2,-1){25}}
\put(80,30){\line(2,1){25}}
\put(80,30){\line(2,-1){25}}
\multiput(40,30)(11,0){4}{\line(1,0){7}}
\put(40,30){\makebox(0,0){$\times$}}
\put(12,46){\makebox(0,0){$\nu^c$}}
\put(10,15){\makebox(0,0){$\nu$}}
\put(110,45){\makebox(0,0){$u$}}
\put(111,16){\makebox(0,0){$u^c$}}
\put(60,10){\makebox(0,0){(a)}}
\end{picture}}
\put(140,0){\begin{picture}(80,90)
\put(40,70){\line(-2,1){25}}
\put(40,70){\line(2,1){25}}
\put(40,30){\line(-2,-1){25}}
\put(40,30){\line(2,-1){25}}
\multiput(40,30)(0,11){4}{\line(0,1){7}}
\put(40,70){\makebox(0,0){$\times$}}
\put(12,86){\makebox(0,0){$\nu^c$}}
\put(10,16){\makebox(0,0){$\bar u$}}
\put(70,85){\makebox(0,0){$\bar\nu$}}
\put(71,16){\makebox(0,0){$u^c$}}
\put(40,10){\makebox(0,0){(b)}}
\end{picture}}
\put(230,20){\begin{picture}(90,55)
\put(40,30){\line(-1,0){30}}
\put(40,30){\line(2,1){36}}
\multiput(40,30)(12,-6){3}{\line(2,-1){10}}
\put(40,30){\makebox(0,0){$\times$}}
\put(6,32){\makebox(0,0){$\nu^c$}}
\put(80,50){\makebox(0,0){$\bar\nu$}}
\put(80,10){\makebox(0,0){$\Phi$}}
\put(40,10){\makebox(0,0){(c)}}
\end{picture}}
\put(330,0){\begin{picture}(80,90)
\put(40,70){\line(-2,1){25}}
\put(40,70){\line(2,1){25}}
\put(40,30){\line(-2,-1){25}}
\put(40,30){\line(2,-1){25}}
\multiput(40,30)(0,11){4}{\line(0,1){7}}
\put(40,30){\makebox(0,0){$\times$}}
\put(40,70){\makebox(0,0){$\times$}}
\put(11,85){\makebox(0,0){$\nu^c$}}
\put(11,19){\makebox(0,0){$\nu^c$}}
\put(70,85){\makebox(0,0){$\bar\nu$}}
\put(70,17){\makebox(0,0){$\nu$}}
\put(40,10){\makebox(0,0){(d)}}
\end{picture}}
\end{picture}
\caption{Feynman diagrams contributing to plasma scattering (a,b),
massive bound state decay (c) and vorton decay (d).}
\label{Feynman fig}
\end{figure}
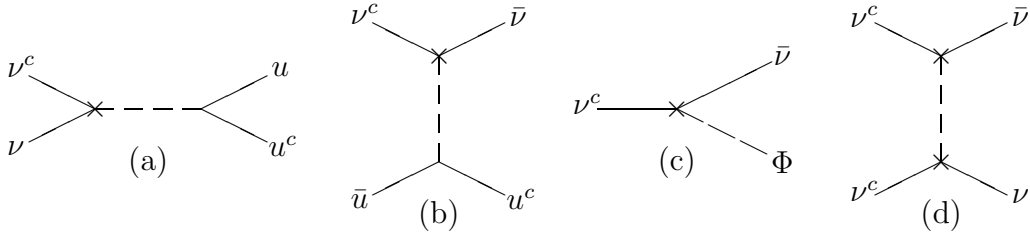

Having established the spectrum of fermion states in an abelian
string background, we now turn our attention to the stability of
charge carriers on these strings. In the absence of other particles,
currents carried by zero modes on isolated, straight strings are
stable on grounds of energy and momentum conservation. However, in a
realistic setting there are many processes which can depopulate zero
modes. Strings are not isolated and in the early Universe they and
their bound states will interact with the hot plasma. Also, bound
states on different strings, or different parts of a single curved
string, can scatter from one another. If there are couplings within
the theory that allow a heavy neutrino zero mode to scatter from a
light plasma particle to produce a light fermion-antifermion pair via
an intermediate electroweak Higgs boson, we can have charge carrier decay
from the Feynman diagrams in figures~\ref{Feynman fig}a and
\ref{Feynman fig}b. Also of interest is the decay of massive bound
modes into light fermions (see figure~\ref{Feynman fig}c). In theories
which allow such interactions, the charge carriers can decay. It is
also possible for heavy neutrino currents on different strings to
decay by exchanging a Higgs particle, as in figure~\ref{Feynman fig}d.
The aim of this section is to calculate these decay rates.

The details of the calculations are given in the appendices, here we
highlight the important differences between the calculation in the
background of the string and the corresponding trivial background
case. We initially consider our system to be restricted to some box
of finite volume, $V$. Following the canonical procedure we decompose
the field operators into a sum over orthonormal wavefunctions and
creation/annihilation operators. Throughout we normalise our
wavefunctions according to,
\be
\int d^3x\, \phi^\dagger_\bk(\bx)\phi_{\bk'}(\bx)=\delta_{\bk\bk'} \ .
\ee
We take the following decomposition for scalar fields,
\be
\hat\Phi(t,{\bf x})=\sum_{\bf k}
{1\over \sqrt{2w}}\bigl(\hat a_{\bf k}\phi_{\bf k}({\bf x})e^{-iwt}
       +\hat b^\dagger_{\bf k}\phi_{-\bf k}({\bf x})e^{iwt}\bigl) \ ,
\label{phiexp}
\ee
where $w^2={\bf k}\cdot{\bf k}+m_{\phi,k}^2$. The necessary commutation
relations are, $ [\hat a_{\bf k},\hat a^\dagger_{\bf k'}] = 
[\hat b_{\bf k},\hat b^\dagger_{\bf k'}]= \delta_{{\bf k}{\bf k'}}$.

The corresponding decomposition of fermionic fields is,
\be
\hat\Psi(t,\bx)=\sum_\bk \left(\hat c_\bk U_\bk(\bx) e^{-iwt}
                 + \hat d^\dagger_\bk V_\bk(\bx) e^{iwt}\right) \ ,
\label{psiexp}
\ee
where the states $U_\bk$ and $V_\bk$ are spinor valued
and the sum is over momentum states. In this case we
impose anticommutation relations $\{\hat c_\bk,\hat c_{\bk'}^\dagger\}
=\{\hat d_\bk,\hat d_{\bk'}^\dagger\}=\delta_{\bk\bk'}$.

With these normalisations we have a simple interpretation of the amplitude,
\be
{\cal A}=\bra{1,2} S \ket{3,4} \ .
\ee
The probability of the interaction, characterised by $S$, converting the
initial state, $\ket{1,2}$, into the final state, $\ket{3,4}$, is simply,
\be
P=\vert {\cal A}\vert^2 \ .
\ee
We can now consider a simple, second order tree diagram in
$\Phi \Psi^\dagger \Psi^c$ theory, where the interaction is given by,
\be
{\cal H}_{\rm int} = ig\Phi \Psi_1^\dagger i\sig{2}\Psi_2^* 
+ ig^*\Phi^* \Psi_3^T i\sig{2}\Psi_4 + \hconj \ .
\ee
The amplitude is given by,
\be
{\cal A}=\bra{\mbox{in}} T |g|^2 i^2
\int d^4x \Psi^\dagger_1(x) i\sig{2} \Psi^*_2(x) \Phi_I(x)
\int d^4y \Psi^T_3(y) i\sig{2} \Psi_4(y) \Phi^*_I(y)
\ket{\mbox{out}} \ ,
\ee
where incoming and outgoing states have the form,
\be
\bra{\mbox{in}} = \bra{0} \hat c_1 \hat c_2 \ , \quad 
\ket{\mbox{out}} = \hat c_4^\dagger \hat c_3^\dagger \ket{0} \ .
\ee
Expanding the field operators in the usual way we obtain,
\bea
&&{\cal A} = |g|^2 \sum_k \int d^4x \int d^4x' \, 
 U^\dagger_1(\bx) i\sig{2} U^*_2(\bx) e^{i(w_1 +w_2)t} 
\nonumber \\ && \hspace{1in}
\times \bra{0}T \Phi_I(x) \Phi_I(y)\ket{0}
U^T_3(\bx') i\sig{2} U_4(\bx') e^{-i(w_3+w_4)t'} \ .
\eea
Expressing the Green's function as a sum over a complete set of states,
we have,
\bea
&&{\cal A}=|g|^2 \sum_k \int d^4x \int d^4x' \, 
 U^\dagger_1(\bx) i\sig{2} U^*_2(\bx) e^{i(w_1 +w_2-w_I)t} 
\nonumber \\ && \hspace{1in}
\times \frac{ \phi_\bk(\bx)\phi^*_\bk(\bx')}{{\cal T}
[w_I^2-\bk_I^2-m_{\phi,k}^2]} 
U^T_3(\bx') i\sig{2} U_4(\bx') e^{-i(w_3+w_4-w_I)t'} \ ,
\eea
where ${\cal T}$ is the temporal extent of the region we are considering
and $m_{\phi,k}$ is the (constant) effective mass of the scalar mode.

It is useful to consider briefly how this calculation would proceed
in a trivial background. In a space-time box of volume $V{\cal T}$,
we have,
\be
\sum_k \to {V{\cal T}\over (2\pi)^4}\int d^4k_I \ .
\ee

Each wavefunction introduces a factor of $V^{-1/2}$ to the
amplitude. The spatial integrations yield finite volume
approximations to energy and momentum conserving delta functions which
we denote $\delta^{4}_{{\cal T}V}$, where 
$\int e^{ik\cdot x} d^4x=(2\pi)^4 \delta^{4}_{{\cal T}V}(k)$ and 
$\delta^{4}_{{\cal T}V}(0) = {\cal T}V(2\pi)^{-4}$.

The value of the propagator is fixed by energy/momentum
conservation and, up to dimensionless factors, we have,
\be
{\cal A} \sim {|g|^2 \over {V^2}}
{1\over k_I^2-m_\phi^2}
\delta^{4}_{{\cal T}V} (k_1 + k_2 - k_3 - k_4) \ ,
\ee
where $k_I = k_1+k_2 = k_3+k_4$.

The interaction cross section is given by,
\be
\sigma=\sum_{\rm final \ states} \frac{V|{\cal A}|^2}{{\cal T}v_{\rm rel}} \ .
\ee 
Replacing this sum by integrals over the final state momenta and
including all the spinor factors, we find,
\be
\sigma={\vert g\vert^4 \over 16\pi} { {k_I^2}\over (k_I^2-m_\phi^2)^2} \ .
\ee
Now we can consider a situation of interest.  If the incoming
particles are a zero mass particle of energy $w_{zm}$ and a light
plasma particle of energy of order the temperature, then
$k_I^2 \sim w_{zm}T$ and,
\be
\sigma\sim{\vert g\vert^4 w_{zm}T \over (w_{zm}T-m_\phi^2)^2} \ . 
\ee

Having discussed the calculation in a trivial background, we can now
consider the string background. The major difference arises in the
spatial integration associated with the position of the initial
vertex. Let particle 1 be the bound state,
\be
\Psi_1(x)\sim e^{iw_1 t-ik_{1z}z-in_1\theta}{\mt \over \sqrt{L}} e^{-\mt r} \ ,
\ee
where $\mt^{-1}$ is the radius of the bound state.  As this state is
localised close to the string, it will only overlap significantly with
the lowest angular mode components of the incoming scattering wave and
so the integration over the position of the initial vertex will be
dominated by these modes (and for the same reason the lowest modes of
the propagator).

We consider the incoming particle to be a plane wave asymptotically
and do a mode by mode matching of this incoming plane wave onto the
states in the string background. As we only have square integrable
wavefunctions in the string background, in general we must also
include an outgoing scattered wave: this is the source of
Aharanov-Bohm scattering.

The general features of both the incoming particle and intermediate
particle wavefunctions are a growth region for $k_T r \alt 1$ and an
oscillatory region for $k_T r \agt 1$, where $k_T$ is the momentum of the
particle transverse to the string \cite{wfnamp}. The transverse
momentum of the incoming particle will be much less than $\mt$, so we
expect some power law type behaviour in the region of overlap with the
bound state. Similarly, if the transverse momentum of the intermediate
particle, $k_{IT}$ is smaller than $\mt$, its wavefunction will also be
roughly a power law in the important overlap region. The integration
over the initial vertex position then produces a number that is
suppressed due to the limited extent of the bound state wavefunction,
but is only mildly dependent on $k_{IT}$. Only when $k_{IT}$ exceeds
$\mt$ do we find a wavefunction oscillating in the overlap region. Thus
only for $k_{IT} \agt \mt$ does the integral decrease. Instead of a
momentum conserving delta function, this spatial integration yields
something more like a step function, which permits transverse momentum
non-conservation up to the mass scale $\mt$. This behaviour is natural
as the string breaks transverse translation invariance and the object
bound to the string has an effective radius $\mt^{-1}$.

The violation of momentum conservation in the string background opens
up significant areas of phase space that are forbidden in the trivial
background. Of particular importance is the possibility of resonant
scattering. We consider first the decay of a charge carrier into a
fermion and a Higgs boson. In the string background transverse momentum can
be {\it acquired} from the string to place the Higgs particle on shell
in a much larger region of phase space. This introduces a factor of
$\mt^2\mB$. The small size of the region in which the wavefunctions
overlap gives an extra factor of $\mt^{-2}$. The full decay into three
fermions can then be considered to be an initial decay into a light
fermion and physical electroweak Higgs boson, followed by Higgs decay. In
appendix~\ref{ap bs} it is shown that the massive bound state lifetime is
\be
\tau \sim (|g_\nu|^2 \mB)^{-1} \ ,
\label{bound life}
\ee
where $g_\nu$ is the Yukawa coupling in the neutrino electroweak mass term. 
Thus \bref{bound life} results in a small lifetime.

In arriving at this lifetime we have made certain assumptions, in
particular we have neglected the possibility that the incoming and
outgoing fermionic wavefunctions may be amplified near the string
\cite{wfnamp}. We have also neglected back-reaction on the string. As
a significant contribution to the amplitude comes from large
transverse momentum non-conservation, the use of a static string
background might not be a good approximation. However, if the string
recoils it will absorb some of the energy of the interaction. This
will reduce the amount of momentum non-conservation required to make
the outgoing particles on-shell. Thus these approximations are not likely to
increase the lifetime, and \bref{bound life} should therefore be taken
as an overestimate.

The fate of massless bound states is also complicated by momentum
non-conservation. Of particular interest are the bound states that
stabilise vortons. Contraction of the vorton will ensure that the
states at the Fermi surface will have GUT scale momenta. The lifetime
of these high momentum states is critical, if they decay the vorton
will contract and {\it promote} low momentum states to high
momentum. Energy and $z$-momentum conservation prevents the massless
states decaying spontaneously, in contrast to the massive bound
states. It is however possible for them to decay by interaction with
plasma particles or other zero modes. If $\ms > M$ it is also possible
for high energy massless currents on a curved string to decay by
tunneling to free heavy neutrinos. However the rate will
not be significant unless $\ms \gg M$~\cite{vorton}.

If we consider one of these states scattering from a typical
plasma particle, we have a centre of mass energy of order
$\sqrt{\ms T}$, well above the mass of the intermediate
particle. Transverse momentum non-conservation again allows for
resonant scattering. Including amplification of the incoming plasma
particle wavefunction, the lifetime of these high momentum zero modes
is found to be,
\be
\tau \sim {\mt^2\over |g_\nu|^2 T^3}\left({T\over \mt}\right)^{2Q} \ ,
\ee
where $Q$ is the charge of the plasma particle under the string gauge
field. We have taken the plasma particles to be massless and ignored
any temperature dependent corrections to their mass. Providing 
$\sqrt{w_{zm}T} > m_\phi \sim T$, resonant scattering is possible and
the cross-section is largely independent of the zero mode energy,
$w_{zm}$. The unamplified lifetime of the modes thus scales with the
plasma density. Conversely, the amplification factor decreases with
increasing temperature as the ratio of the GUT scale to the typical
thermal energy grows.

In the radiation dominated era we can take $t=\alpha T^{-2}$ and the
lifetime becomes,
\be
\tau \sim \mt^{2-2Q} |g_\nu|^{-2} T^{2Q-3}=\mt^{2-2Q} |g_\nu|^{-2} 
\alpha^{Q-3/2}t^{3/2-Q} \ .
\ee 
The probability of a zero mode state scattering in time interval $dt$
is $dt/\tau$, thus the probability of a zero mode state scattering
after some time $t_i$ is,
\be
P({\rm decay}\hskip 3pt {\rm after} \hskip 3pt t_i)
= 1 - e^{-\int_{t_i}^\infty {dt\over \tau}}
= 1 - \exp\left\{-\mt^{2Q-2} |g_\nu|^2 \alpha^{3/2-Q}{t_i^{Q-1/2}\over 1/2-Q}
\right\} \ .
\ee
If the magnitude of the exponent is small there is a small probability,
thus zero modes are stable if,
\be
t_i > O\left(
\left[\mt^{2Q-2} |g_\nu|^{2} \alpha^{3/2-Q}\right]^{2/(1-2Q)}\right) \ .
\ee
Now, $\alpha \sim M_{Pl}/10$ and in the case of SO(10), $Q=3/10$, leading to
the condition,
\be
t_i > O\left([\mt^{-7} |g_\nu|^{10} \alpha^{6}]\right) > 
O\left(\left[{M_{Pl}\over 10\mt}\right]^6 |g_\nu|^{10}\mt^{-1}\right) \ .
\ee  

As the lifetime varies only slightly faster than $T^{-2}$, this result
for $t_i$ is very sensitive to the Yukawa coupling. For $\mt \sim
10^{15}$GeV, if $|g_\nu|=1$ zero mode states populated after $t_i\sim
10^{15}t_\sGUT$ will be stable, while if $|g_\nu| \alt 0.03$, this
scattering is never significant. In the SO(10) model $g_\nu$ is also
the Yukawa coupling for the corresponding quarks, thus there is an
epoch when $\nu^c_\tau$ zero modes will scatter from the string, but
$\nu^c_e$ and $\nu^c_\mu$ zero modes will never scatter by this
process. Thus the interaction with plasma particles can not significantly
remove zero modes from the string.  
Note that it is also possible to create currents using the
above interactions in reverse. Hence, if thermal equilibrium is reached the
number density of zero modes will be of order $T$.

Within the SO(10) model there is also the possibility of mediating
these processes by GUT mass Higgs fields with zero VEV. In this case
the Yukawa coupling need not be small, but the centre of mass energy
of the interaction is only of order the intermediate particle mass for
$T\sim T_\sGUT$. Thus below the GUT temperature the reaction rates for
these processes are rapidly suppressed by powers of $T/T_\sGUT$.

None of these plasma scattering processes can remove $\nu^c_e$ and
$\nu^c_\mu$ zero modes, and are only significant for $\nu^c_\tau$ immediately
after the phase transition. Thus, they are unable to prevent the vorton
density from dominating the energy density of the Universe.

The plasma scattering processes considered above failed to remove zero
modes due to the decreasing plasma density at late times. A distinct
category of process is the scattering of a zero mode on one string by
a second zero mode on another string. This is particularly relevant
for vortons as they form small loops with a typical radius only one or two
orders of magnitude larger than the string width. We thus have zero
modes moving in opposite directions on opposite sides of the
vorton. For simplicity the decay rate can be calculated by considering
two straight, anti-parallel strings (appendix~\ref{ap vort}). Physically one
expects a suppression due to the finite separation of the initial and
final vertices. As the initial and final vertices are confined to
different pieces of string, the amplitude contains a factor of 
$e^{i k_TR}$, where $k_T$ is the transverse momentum of the intermediate
particle and $2R$ is the separation of the strings. When the integration
over the intermediate momenta is performed, this factor can cause the
integrand to oscillate, leading to a suppression of the
amplitude. This occurs if $k_TR$ is large where the standard
propagator factor is significant.

If the intermediate particle is an electroweak Higgs boson, then the standard 
propagator factor
peaks for $k_T$ around the electroweak scale, $k_TR$ is tiny in the
important region, there is no oscillation and no suppression. The
cross-section in this case is found to be,
\be
\sigma\sim {|g_\nu|^4\over (M_\sGUT R)^4} \ .
\label{vor sigx}
\ee
This cross-section is dimensionless as the scattering is effectively
in one spatial  dimension. In this case resonant scattering is not
possible, thus the Higgs width does not enter and the rate therefore
contains a factor of $g_\nu^4$.
 
Conversely, for a GUT scale mass as the intermediate particle, the standard 
propagator
factor peaks for $k_T\sim M_\sGUT$, giving $k_T R\sim 10-100$. In this
regime the reaction rate is found to display the exponential
suppression expected on physical grounds,
\be
\sigma\sim {|g_\nu|^4 \over (M_\sGUT R)^3} e^{-4 M_\sGUT R} \ .
\label{vor sigG}
\ee
The exponential suppression in \bref{vor sigG} makes such processes
irrelevant in all  physical situations.

Taken at face value, if electroweak particles mediate current--current
scattering on different segments of string then \bref{vor sigx} gives
a short lifetime for charge carriers on a vorton. However, there are
complications in directly applying this result derived for straight
strings to curved vortons. The presence of charge carriers on the
vorton loops results in the loop carrying angular momentum, which must
be conserved in any physical process under investigation. The above
calculation does not take into account conservation of angular
momentum, as the system considered has no rotational symmetry.
However, the angular momentum gives rise to a centrigual barrier,
suppressing the above decay channel. Combined with energy
conservation, this would prevent massless modes on the string
scattering into massive modes. However, the fermionic spectrum has not
been calculated for a circular loop and there is no reason to expect
the zero modes to remain massless and thus, angular momentum
conservation can only be considered in a consistent rotationally
invariant calculation. The calculation above works consistently with
straight strings, thus while \bref{vor sigx} may not be directly
applicable to vorton decay, it is relevant for interactions on
non-circular loops and scattering of currents on a string network.

At very early times in the friction dominated regime, the string
correlation length is small. The interaction of \bref{vor sigx} could
be operative. However the string density drops too quickly for it to
be significant. As the density of current carriers builds up, the
inter-string separation increases, so the cross-section decreases
significantly. Consequently, the current-current interaction is unable
to reduce the number of current carriers on the string and prevent
vorton formation.

\section{Summary}
\label{sum}

In this paper the existence and form of fermion bound states and their
corresponding currents on cosmic strings were investigated. Only
time-like and light-like fermions can exist on the string, in contrast to
the bosonic case. Using numerical 
methods,
the discrete spectrum of states for an abelian string model with one
fermion field was determined. We found that the number of bound states
increased as the Yukawa coupling grew. For very low values there are
just two massive bound states and a zero mode.

Bound mode states will always occur. The occupancy of
these states depends on the decay modes of the carriers. Since the
cosmic string breaks Lorentz invariance, transverse momentum is not
conserved in interactions in a string background. This leads to an
enlarged phase space and increased cross
sections. On the other hand, the fact that the bound states are
confined to the string reduces the overlap of the particle
wavefunctions, tending to reduce the cross sections.

Unless the massive carriers
are isolated from the electroweak sector, they will decay and the states
will empty on a time scale of $10^{-33}$s. If the carriers are stable,
they will persist on the string and carry angular momentum, contributing
to vorton formation.

In order for the massless current carriers to decay they must interact
with other particles, such as carriers on other strings or plasma
particles.  The most significant decays involve electroweak Higgs 
intermediate states. Non-conservation of momentum means that plasma
scattering is usually resonant. However the small size of the Yukawa
couplings and the plasma density mean that the rate of this decay is
too small to be cosmologically significant (at least for $\nu_{e R}$
and $\nu_{\mu R}$ current carriers). Decays involving GUT boson
intermediates are also possible, although the momentum
non-conservation is not large enough to make these decays
resonant. The interaction rate in this case is also small.

Massless current carriers can also decay by scattering with currents
on other strings. The cross section for this interaction is tiny
unless the two strings are very close together. For GUT mass
intermediate states appropriate densities are never physically
realised, but for electroweak intermediate states suitable densities
arise immediately after string formation. A situation similar to this
occurs when the current is on a string loop. In this case the current
will decay rapidly. Unfortunately this result cannot be applied
directly to circular loops, where angular momentum conservation must
be taken into account. Consequently, it can not be applied directly to
the case of cosmic vortons. This situation requires more detailed
analysis~\cite{in prep}.

In the SO(10) model, after the electroweak phase transition, the
situation changes dramatically. The right-handed neutrino zero modes
mix with left-handed neutrinos allowing currents on curved strings to
decay by tunneling into free left-handed neutrinos. This averts any
cosmological disaster.

However, if the zero mode is isolated from the electroweak sector,
then the decay processes considered in this paper are not
operative. As a consequence, the zero mode survives. For GUT scale
strings, this results in a cosmological disaster as discussed in
\cite{brandon&acd}.

\acknowledgements

This work was supported in part by PPARC, the ESF and the EU under TMR
grant no.\ ERBFMRXCT97--0122. We are also grateful to the CNRS--Royal
Society exchange programme for support. We wish to thank Brandon
Carter, Patrick Peter and Neil Turok for discussions.

\appendix
\section{Basics: Propagators, Interactions and Wavefunctions}
\label{ap conv}

\subsection{The Scalar Propagator}

We will begin by finding the propagator, $G(x,y)$, which satisfies,
\be
\hat O G(x,y)=(\partial_t^2-\nabla^2+V(r)^2)G(x,y)=-i\delta^4(x-y),
\ee
where $\hat O\Phi=0$ is the equation of motion. $V(r)^2$ represents
the position dependent potential that the scalar field experiences in
the string background. The standard connection with the time ordered
product holds here too;
\be
\hat O \bra{0}T\Phi(x)\Phi^*(x') \ket{0}=-i\delta^4(x-y).
\ee
We now have to find a suitable representation of this Green's function.
Consider the quantity,
\be
G(x-x')=
i\sum_k \frac{ e^{-iw\tau} }{{\cal T}[w^2-\bk^2-m_{\phi,k}^2]}
\phi_\bk(\bx)\phi^*_\bk({\bf x'}),
\label{greens}
\ee
where $\tau=t-t'$ and the sum extends over states of all masses,
i.e.\ we include off-shell states in the sum. We will work in a
space-time box and so impose boundary conditions at $t=-{\cal T}/2$ and
$t={\cal T}/2$. The allowed energies are then quantised in just the same way
as the momenta. Normalising these states to unity introduces a factor
of $1/\sqrt{\cal T}$ into each wavefunction. We will display this factor
explicitly so that only the $1/\sqrt{V}$ factors are implicit in any
wavefunction. 
 
The full wavefunctions, $\phi_k(x)=e^{-iwt}\phi_\bk(\bx)$, satisfy,
\bea
\hat O \phi_k(x) &=&[-w^2-\nabla^2+V(r)^2] e^{-iwt}\phi_\bk(\bx) 
\nonumber \\
&=&[-w^2+\bk^2+m^2_{\phi,k}] e^{-iwt}\phi_\bk(\bx) \ ,
\eea
where $m_{\phi,k}$ is the (constant) effective mass of the scalar mode
of momentum $k$.

To see that we have a Green's function, act with the operator on $G$
and use the fact that the off shell states are a complete set of
solutions of the 4-dimensional eigenvalue problem. We tame the
singularity in the usual way with an $i\epsilon$ to produce the
Feynman propagator.

\subsection{The Interaction}

Now consider the simple, second order tree diagram in a theory with
\be
{\cal H}_{\rm int} = ig\Phi \Psi_1^\dagger i\sig{2}\Psi_2^* 
+ ig^*\Phi^* \Psi_3^T i\sig{2}\Psi_4 + \hconj \ .
\ee
We will start by considering incoming and outgoing states of the form
\be
\bra{\mbox{in}} = \bra{0} \hat c_1 \hat c_2 \ , \quad 
\ket{\mbox{out}} = \hat c_4^\dagger \hat c_3^\dagger \ket{0} \ .
\ee
Substituting in the expressions for the external fields \bref{psiexp} we
arrive at the VEV of the time ordered product in the usual way,
\bea
&&{\cal A}=|g|^2 \sum_k \int d^4x \int d^4x' \, 
 U^\dagger_1(\bx) i\sig{2} U^*_2(\bx) e^{i(w_1 +w_2-w_I)t} 
\nonumber \\ && \hspace{1in}
\times \frac{ \phi_\bk(\bx)\phi^*_\bk(\bx')}{{\cal T}[k_I^2-m_{\phi,k}^2]} 
U^T_3(\bx') i\sig{2} U_4(\bx') e^{-i(w_3+w_4-w_I)t'} \ .
\label{free amp2}
\eea

\subsection{The Amplitude}

First we will evaluate the amplitude \bref{free amp2} in a
trivial background, using cartesian coordinates. We will be mainly
interested in 2-component left-handed massless Weyl fermions, with
Lagrangian ${\cal L}=\Psi^\dagger i \sig{\mu} D_\mu \Psi$. The wavefunctions
in \bref{phiexp} and \bref{psiexp} are then
\bea
&&\phi_\bk(\bx) = \frac{e^{i\bk \cdot \bx}}{\sqrt{V}} \nonumber \\ &&
U_\bk(\bx) = \frac{e^{i\bk \cdot \bx}}{\sqrt{V}} u_\bk\ \ \ \
V_\bk(\bx) = \frac{e^{-i\bk \cdot \bx}}{\sqrt{V}} e^{-i\beta} u_\bk
 \nonumber \\ &&
\mbox{with} \ \ u_\bk = \frac{1}{\sqrt{2w}}
\bmat{c} \sqrt{w+k_z} \\ \sqrt{w-k_z} e^{i\beta} \emat \ ,
\label{str states}
\eea
where $k_x + ik_y = k_T e^{i\beta}$. Since we are considering massless
fermions $w = |\bk|$. The sum over intermediate states can be approximated
by an integral,
\be
\sum_k \to {V{\cal T}\over (2\pi)^4}\int d^4k_I \ .
\ee
The amplitude then becomes,
\be
{\cal A}=|g|^2 {1\over (2\pi)^4V^2}\int d^4k_I d^4x d^4x'
u_{\bk_1}^\dagger i\sig{2} u_{\bk_2}^*
\frac{ e^{ix\cdot(k_1+k_2-k_I)} e^{-ix'\cdot(k_3+k_4-k_I)}}
{k_I^2-m_\phi^2}
u_{\bk_3}^T i\sig{2} u_{\bk_4} \ .
\label{free amp3}
\ee
We will define $(2\pi)^4 \delta^{4}_{{\cal T}V}(k)=\int e^{ik\cdot x} d^4x$
to be the equivalents of the usual delta functions at finite time and volume,
with $\delta^{4}_{{\cal T}V}(0) = {\cal T}V(2\pi)^{-4}$. The large time and
volume limits are implied.
 
Evaluating the integrals gives
\be
{\cal A} = \frac{|g|^2 (2\pi)^4}{V^2}
\frac{u_{\bk_1}^\dagger i\sig{2} u_{\bk_2}^* u_{\bk_3}^T i\sig{2} u_{\bk_4}}
{k_I^2-m_\phi^2}
\delta^{4}_{{\cal T}V} (k_1 + k_2 - k_3 - k_4) \ ,
\label{free amp4}
\ee
with $k_I = k_1+k_2 = k_3+k_4$.

The total cross section for the interaction is obtained by summing over the
possible final states
\be
\sigma=\sum_{\rm final \ states} \frac{V|{\cal A}|^2}{{\cal T}v_{\rm rel}} \ .
\ee
The sum over final states can be replaced by integrals over the
final state momenta,
\be
\sigma= \frac{V}{{\cal T}v_{\rm rel}}
\int \frac{V d^3k_3}{(2\pi)^3} \int \frac{V d^3k_4}{(2\pi)^3}
\frac{|g|^4 (2\pi)^8}{ V^4} \delta^{4}_{{\cal T}V}(0)
\delta^{4}_{{\cal T}V} (k_1 + k_2 - k_3 - k_4)
\frac{|u_{\bk_1}^T i\sig{2} u_{\bk_2}|^2 |u_{\bk_3}^T i\sig{2} u_{\bk_4}|^2}
{(k_I^2-m_\phi^2)^2} \ .
\label{free prob1}
\ee

The relative velocity of the incoming particles is
$v_{\rm rel} = k_1^\mu k_{2\mu}/(w_1 w_2)$. Using this and
$|u_\bk^T i\sig{2} u_{\bk'}|^2 = k^\mu k'_{\mu}/(2w w')=(k+k')^2/(4w w')$
gives
\be
\sigma=\frac{|g|^4}{8(2\pi)^2} 
\frac{k_I^2}{(k_I^2-m_\phi^2)^2}
\int \frac{d^3k_3}{w_3} \frac{d^3k_4}{w_4}
\delta^{4} (k_3 + k_4 - k_I) \ .
\label{free prob2}
\ee
The integral is Lorentz invariant, and can easily be evaluated in the centre
of momentum frame to give
\be
\sigma=\frac{|g|^4}{16\pi} \frac{k_I^2}{(k_I^2-m_\phi^2)^2}
\label{free prob3}
\ee

\subsection{The Higgs Width}

The part of the electroweak Higgs which couples to neutrinos can decay
into two quarks, $\Phi \rightarrow q + q^c$, where $q=u, c, t$. This
is a first order process with rate given by
\bea
\Gamma_\phi &=& 3\sum_{q=u,c,t} 
\int \frac{V^2 d^3k_1 d^3k_2}{{\cal T}(2\pi)^6}
\left| ig_q^*(2\pi)^4 \frac{u_{\bk_1}^T i\sig{2} u_{\bk_2}}{\sqrt{2m_\phi V^3}}
\delta(m_\phi - w_1 - w_2)\delta^3(\bk_1 + \bk_2) \right|^2 \nonumber \\ 
&=& 3\sum_{q=u,c,t} \frac{|g_q|^2 m_\phi}{8(2\pi)^2}
\int \frac{d^3k_1}{w_1}\frac{d^3k_2}{w_2}
\delta(m_\phi - w_1 - w_2)\delta^3(\bk_1 + \bk_2)
\nonumber \\
&=& 3\sum_{q=u,c,t} |g_q|^2 \frac{m_\phi}{16\pi} \ .
\label{higgs width}
\eea

\subsection{Wavefunctions in a Cosmic String Background}

Having established our normalisations and conventions in 
the familiar trivial background, we now turn to the
calculation in the background of a cosmic string. 
We work in a cylindrical box of length $L$ and volume $V$.
The normalised on-shell states are given in section~\ref{spectrum}.

In the string background it is natural to perform an expansion
in angular mode number, yet we still wish to consider an incoming
plane wave. The two expansions can be matched by first expanding
the plane wave in terms of Bessel functions;
\be 
e^{ik(x\cos{\beta}+y\sin{\beta})} = e^{i(k_xx+k_yy)}
=\sum_{q=-\infty}^{\infty} i^q J_q(kr) e^{iq(\theta-\beta)} \ ,
\label{bessel exp}
\ee
and then asymptotically matching these Bessel functions mode by mode
to the modes in the string background. As we only have square
integrable wavefunctions in the string background, in general we must
also include an outgoing scattered wave. This outgoing wave is the
source of Aharanov-Bohm scattering. Using an angular mode
decomposition of the intermediate particle, all the wavefunctions at
the initial vertex are in the form of cylindrical modes.

We must now consider the participating wavefunctions. Let particle 1
be the bound state, 
\be
\Psi_1(x)\sim e^{iw_1t-ik_{1z}z-in_1\theta}{\mt \over \sqrt{L}} e^{-\mt r} \ .
\ee 

As this state is bound to the string, it will only overlap
significantly with the lowest angular mode components of the incoming
scattering wave, and so the integration over the position of the
initial vertex will be dominated by these components. Similarly, we
need only consider the lowest angular modes of the propagator.
 
The modes of a generic fermion wavefunction in the string background
have three important regions. If the transverse momentum is $k_T$,
these regions are $k_Tr \agt 1$, $1/k_T \agt r \agt 1/\mt$ and $r \alt 1/\mt$.
In the large $r$ region, $k_Tr \agt 1$, we have some Bessel function at
large argument. This region dominates the normalisation integral. In
the intermediate region, $1/k_T \agt r \agt 1/\mt$ we have some Bessel
function at small argument. The order of this Bessel function is
shifted by the string gauge field~\cite{wfnamp}. In the lowest
angular momentum modes there is a mixture of Bessel functions of the
first and second type, leading to one component of the spinor varying
like $r^{-Q}$ in this region, where $Q$ is the magnitude of the charge
of the fermion under the string gauge field. Finally, inside the
string, $r \alt 1/\mt $, the spinor components tend to constants as
$r \to 0$. The effect of the intermediate region is thus to amplify
the wavefunctions by $O([\mt/k_T]^Q)$. In the following appendices we
calculate unamplified cross-sections. The amplification factors,
which depend on the fermion charges, will be added at the end.

For the outgoing states a construction as above is less intuitive.  We
can think of the incoming wave undergoing classical scattering and
some small part of it also partaking in the quantum scattering. The
corresponding interpretation for the outgoing wave is that the quantum
scattering excites states consisting of an outgoing plane wave and
outgoing Aharanov-Bohm scattered waves. Whilst this does not lend
itself to a clean interpretation in terms of two-to-two scattering, it
is just another manifestation of the inappropriateness of the plane
wave as the asymptotic state in the background of the string. The
$1/r$ fall off of the gauge field of the string introduces  long range
interactions in the string background which we will neglect.  This
corresponds to assuming that the outgoing states are free.

The outgoing states we are interested in are light and have
little interaction with the string. Thus we make the
approximation that the outgoing states do not interact with the string
fields and the plane wave expansion employed for a free plane wave can
be used for the outgoing states.

\section{Scattering From a Massless Bound State on a String}
\label{ap zm}

We now consider the problem of real physical interest: a plasma
fermion scattering from a bound fermion via an intermediate scalar to
produce two light fermions. We consider theories with couplings of
the form:
\be
{\cal H}_{\rm int}=ig_\nu\Phi \Psi_{\! \nu^c}^\dagger i\sig{2}\Psi_{\! \nu}^*
+ig_u\Phi \Psi_{\! u}^\dagger i\sig{2}\Psi_{\! u^c}^* + \hconj \ ,
\ee 
where $\Psi_{\nu^c}$ is the heavy neutrino, $\Phi$ is a light Higgs
field, and the other $\Psi$ are light fermions. In the SO(10) model
the couplings $g_\nu$ and $g_u$ are the same.

For the SO(10) model the scattering of interest is
$\nu^c_{\! str} + \nu \rightarrow u + u^c$, where $\nu^c_{\! str}$ is the
massless current carrier. Figure~\ref{Feynman fig}a shows the corresponding
Feynman diagram. The scattering amplitude is then
\bea
{\cal A}&=&\bra{\nu^c_{\! str}(\bk_1) \, \nu(\bk_2)} S 
\ket{u(\bk_3) \, u^c(\bk_4)} \nonumber \\
&=& g_\nu g_u^* \sum_\bk\int d^4x \int d^4y \,
U_{0\bk_1}^\dagger(\bx) i\sig{2} U_{\bk_2}^*(\bx) e^{i(w_1+w_2)t}
\nonumber \\ &&\hspace{1in} \times 
G(x-x') U_{\bk_3}^T(\bx') i\sig{2} U_{\bk_4}(\bx') e^{-i(w_3+w_4)t'} \ .
\label{zm amp1}
\eea
To keep things simple, we will use plane wave approximations of the
Higgs and light fermion states  \bref{str states}. Using these and the
expression for $U_{0\bk}$ \bref{chians0}, the above amplitude can be
expanded to give a similar expression to \bref{free amp3}. Most of the
integrals can be done in the same way as before, leading to
\bea
{\cal A} &=&
g_\nu g_u^* \frac{\mt}{\sqrt{2\pi LV^3}} (2\pi)^2 \int d^2x \,
\delta^{(t,z)} (k_1 + k_2 - k_3 - k_4)
\nonumber \\&&\hspace{1in}
\times \frac{u_{\bk_3}^T i\sig{2} u_{\bk_4}e^{i(\bk_I - \bk_2)\cdot \bx^{(2)}}}
{k_I^2-m_\phi^2+i m_\phi \Gamma_\phi} 
\sqrt{1-\frac{k_{z2}}{w_2}} \, e^{-i\beta_2}
e^{-\int^r M f+\frac{a}{2r'} dr'}\ ,
\label{zm amp2}
\eea
where $\bk\cdot \bx^{(2)} = k_x x + k_y y$. The Higgs particle momentum
satisfies $k_I=k_3+k_4$ and $(k_I)_{z,t}=(k_1 + k_2)_{z,t}$.
The above $d^2x$ integral is most easily evaluated in polar coordinates.
Using a Bessel function expansion of a plane wave \bref{bessel exp} gives
\be
\int e^{i(k_x x+k_y y)} d\theta = 2\pi J_0 (k r)
\ee
Using a numerical solution for the $r$ dependence of the zero mode, we find 
\be
\int J_0 (k r) e^{-\int^r M f+\frac{a}{2r'} dr'} r dr 
\approx \frac{c_0}{\mt^2} e^{-k^2/(2\mt^2)} \ .
\ee
$c_0\approx 0.7$ is a slowly varying function of $M/\ms$. These two results
allow the integrals in \bref{zm amp2} to be evaluated, giving
\bea
{\cal A}
&=& g_\nu g_u^* \frac{(2\pi)^3}{\sqrt{2\pi LV^3}} 
\frac{c_0}{\mt} \sqrt{1-\frac{k_{z2}}{w_2}} \, e^{-i\beta_2}
\delta^{(t,z)} (k_1 + k_2 - k_3 - k_4) 
\nonumber \\&&\hspace{1in} \times
\frac{u_{\bk_3}^T i\sig{2} u_{\bk_4}}
{k_I^2-m_\phi^2+im_\phi \Gamma_\phi} e^{-(k_I-k_2)_T^2/(2\mt^2)}\ .
\eea
The corresponding expression for scattering in a trivial background
has a delta function instead of the gaussian. The physical
significance of this is that transverse momentum is not conserved at
the first vertex. This is to be expected since the string breaks
transverse Lorentz invariance. The extra momentum is absorbed or
provided by the string, up to the scale $\mt=\sqrt{\ms M}/2$. This is
the effective energy scale of the string with respect to bound state
interactions. There is also an extra $\mt^{-1}$ factor, which is a
combination of the bound state normalisation and the fact the
interaction at the first vertex is confined to the string core, whose
effective area is $\mt^{-2}$.

The total cross section is again obtained by squaring the modulus of the
amplitude and integrating over the final state momenta.
After squaring it is convenient to introduce two extra $\delta$-functions
\bea
|{\cal A}|^2 &=& |g_\nu g_u|^2 \frac{\cal T}{V^3} 
\frac{(2\pi)^3 c_0^2}{4\mt^2} \left(1-\frac{k_{z2}}{w_2}\right) 
\frac{k_I^2}{|k_I^2-m_\phi^2+i m_\phi \Gamma_\phi|^2} 
e^{-(k_I-k_2)_T^2/\mt^2} \nonumber \\&&\hspace{1in}
\times \int \frac{d^2k_I}{w_3w_4} \delta^{4}(k_I - k_3 -k_4) \ .
\eea

In this case $v_{\text{rel}} = 1-k_{2z}/w_2$. The $k_3$ and $k_4$
integrals are Lorentz invariant and can be evaluated in the
centre-of-final-momentum frame, as in \bref{free prob2}. Hence
\bea
\sigma&=& \frac{V}{{\cal T}v_{\rm rel}} \int \frac{V d^3k_3}{(2\pi)^3} 
\int \frac{V d^3k_4}{(2\pi)^3}|{\cal A}|^2 \nonumber \\
&=& \frac{|g_\nu g_u|^2}{16\pi^2}
\frac{c_0^2}{\mt^2} \int d^2k_I
\frac{k_I^2\Theta(k_I^2)}
{|k_I^2-m_\phi^2+i m_\phi \Gamma_\phi|^2} 
e^{-(k_I-k_2)_T^2/\mt^2} \ .
\label{zm prob1}
\eea
$\Theta$ is the Heaviside step function.

As $\Gamma_\phi / m_\phi \rightarrow 0$
\be
\frac{k_I^2}{|k_I^2 - m_\phi^2 + i m_\phi \Gamma_\phi|^2}
\rightarrow \frac{\pi m_\phi}{\Gamma_\phi}
\delta(k_I^2 - m_\phi^2) + O(1) \ .
\label{zm prob2}
\ee
If $w_I^2 -k_{Iz}^2 < m_\phi^2$ the first term of \bref{zm prob2} does
not contribute to \bref{zm prob1} and
$\sigma \sim |g_\nu g_u|^2/\mt^2$.  If $w_I^2 -k_{Iz}^2 \ll m_\phi^2$ then
$\sigma \sim |g_\nu g_u|^2 (w_I^2 -k_{Iz}^2)^2/(\mt^2 m_\phi^4)$.
Otherwise, to leading order, the $O(1)$ terms can dropped and
\be
\sigma = |g_\nu g_u|^2 \frac{c_0^2}{16\pi} \frac{1}{\mt^2}
\frac{m_\phi}{\Gamma_\phi}
 e^{-(k_{IT}^2 + k_{2T}^2)/\mt^2} I_0 (2 k_{IT} k_{2T}/\mt^2)
\label{zm sig3}
\ee
with $k_{IT}^2 = w_{\rm in}^2 - k_{{\rm in}z}^2 - m_\phi^2$. The
angular integral is evaluated using
$\int_0^{2\pi} d\theta e^{x\cos\theta} = 2 \pi I_0(x)$. The plasma
particles will have energies of order $T$, so if $w_1$ is less than
the GUT scale then  $k_{IT}, k_{2T} < \mt$ and the exponential terms
can then be dropped.

Scatterings involving charm and top quarks will give similar contributions to
\bref{zm sig3}. Summing these and substituting \bref{higgs width} gives
\be
\sigma = |g_\nu|^2 \frac{c_0^2}{\mt^2} \left(
\Theta(w_{\rm in}^2 - k_{{\rm in}z}^2 - m_\phi^2) + O(|g|^2) + O(T w_1/\mt^2)
 \right) \ .
\label{zm sig4}
\ee

The above calculation only involves the neutral component of the
intermediate Higgs field. Similar scatterings involving the other
components also occur, such as $\nu^c_{\! str} + e \rightarrow d + u^c$. 
In SO(10) there are a total of 5 such scatterings. They all
have similar cross sections to \bref{zm sig4}.

It is also possible for plasma particles to interact with the neutrino
currents by exchanging a virtual Higgs particle, as in
figure~\ref{Feynman fig}b. Conservation of momentum implies that the
Higgs particle is always space-like, and so the scattering is never
resonant. Thus the dominant contributions to the total cross section
come from \bref{zm sig4} and similar interactions.

The rate at which the currents interact with the plasma is given by
$\Gamma_{\rm int} = n_{\rm plas}^{\rm eq} \left< \sigma v_{\rm rel} \right>$,
where $n_{\rm plas}^{\rm eq}$ is the equilibrium number density of each species
of plasma particle and $\left< \sigma v_{\rm rel} \right>$ is the thermally
averaged cross section. $n_{\rm plas} \sim T^3$ and
$\sigma \sim |g_\nu|^2/\mt^2$, so
\be
\Gamma_{\rm int} \sim \frac{|g_\nu|^2}{\mt^2} T^3 \ .
\ee

\section{Decay of a Massive Bound State on a String}
\label{ap bs}

Conservation of energy-momentum prevents massless bound fermions
decaying. The same is not true of massive bound neutrinos. These can
decay into left-handed neutrinos and light Higgs particles 
$\nu^c_{\! str} \rightarrow \bar\nu + \Phi$ (see figure~\ref{Feynman fig}c).
Since they are Majorana fermions, they can also decay into the
corresponding antiparticles $\nu^c_{\! str} \rightarrow \nu + \bar \Phi$.
Both amplitudes involve similar calculations. We will just
consider the lowest mass bound state, with angular dependence $n=-1$.
Typically the heavy neutrino will have a vacuum mass of order the GUT
scale. The mass of a massive bound state will then be some fraction of
the GUT scale.  Using the expressions \bref{chiansM}, \bref{greens},
\bref{str states}, the amplitudes for the above decays in the rest
frame of the bound state are

\bea
{\cal A}_1&=& \bra{\nu^c_{\! str}(\bk_1)} S 
\ket{\bar\nu(\bk_2) \, \Phi(\bk_3)} \nonumber \\
&=& i g_\nu \int d^4x \,
U_{1\bk_1}^\dagger(\bx) i\sig{2} V_{\bk_2}^*(\bx)
\frac{e^{i\bk_3 \cdot \bx}}{\sqrt{2w_3V}}e^{i(w_1-w_2-w_3)t} \nonumber \\
&=& i g_\nu \int d^2x \frac{\mt (2\pi)^2}{\sqrt{8\pi w_3LV^2}}
e^{i(\bk_2 + \bk_3)\cdot \bx^{(2)}} \delta^{(t,z)}(k_1 - k_2 - k_3)
\nonumber \\ &&\hspace{1.5in} \times
\left( \sqrt{1-\frac{k_{z2}}{w_2}} \chi_1 e^{i\theta}
	+ i\sqrt{1+\frac{k_{z2}}{w_2}} \chi_2 e^{i\beta_2} \right) \ , \\
{\cal A}_2&=& \bra{\nu^c_{\! str}(\bk_1)} S 
\ket{\nu(\bk_2) \, \bar \Phi(\bk_3)} \nonumber \\
&=& i g_\nu^* \int d^4x \,
\tilde U_{1\bk_1}^T(\bx) i\sig{2} U_{\bk_2}(\bx)
\frac{e^{i\bk_3 \cdot \bx}}{\sqrt{2w_3V}}e^{i(w_1-w_2-w_3)t} \nonumber \\
&=& i g_\nu^* \int d^2x \frac{\mt (2\pi)^2}{\sqrt{8\pi w_3LV^2}}
e^{i(\bk_2 + \bk_3)\cdot \bx^{(2)}} \delta^{(t,z)}(k_1 - k_2 - k_3) 
\sqrt{1-\frac{k_{z2}}{w_2}}\chi_3 e^{i(\beta_2+\theta)}\ . 
\eea
As with the massless current scattering \bref{zm amp2}, we will use a
Bessel function expansion \bref{bessel exp} of 
$e^{i(\bk_2 + \bk_3)\cdot \bx^{(2)}}$, and approximations of the radial
integrals.
\be
\int J_0 (kr) \chi_2(\ms r) r dr 
\approx \frac{c_2}{\mt^2} e^{-k^2/(2\mt^2)}
\ \ , \ \ 
\int J_1 (kr) \chi_{1,3}(\ms r) r dr \approx \frac{c_{1,3} k}{\mt^2 \mB}
e^{-k^2/(2\mt^2)} \ ,
\ee
with $c_2 \approx -0.7$ and $c_1 \approx c_3 \approx 0.3$. Thus 
\bea
{\cal A}_1 &\approx& -\frac{g_\nu (2\pi)^3}{\mt V\sqrt{8\pi w_3L}}
\left( \sqrt{1-\frac{k_{z2}}{w_2}} c_1 \frac{k_{sx} + ik_{sy}}{\mB}
	+ \sqrt{1+\frac{k_{z2}}{w_2}} c_2 e^{i\beta_2} \right)
\nonumber \\ && \hspace{1.5in} \times
	\delta^{(t,z)}(k_1 - k_2 - k_3) e^{-k_s^2/(2\mt^2)} \ , \\
{\cal A}_2 &\approx& -\frac{g_\nu^* (2\pi)^3}{\mt V\sqrt{8\pi w_3L}}
	 \sqrt{1-\frac{k_{z2}}{w_2}} c_3 \frac{k_{sx} + ik_{sy}}{\mB} 
	e^{i\beta_2} \delta^{(t,z)}(k_1 - k_2 - k_3) e^{-k_s^2/(2\mt^2)} \ ,
\eea
where $(k_s)_{x,y} = (k_2+k_3)_{x,y}$. As with the massless scattering
in appendix~\ref{ap zm}, transverse momentum is not conserved and
there is a $\mt^{-1}$ suppression due to the bound state being confined
to the string core. The 2-component vector $\bk_s$ gives the momentum
contributed by the string. The two amplitudes correspond to different
decay products so the total decay rate is obtained by squaring and
then adding them, and finally summing over the outgoing state
momenta. Before doing this, it is useful to introduce 2 delta
functions to separate out the $\bk_s$ dependence.
\bea
\Gamma &=& \frac{1}{\cal T} \int \frac{V^2 d^3k_2 d^3k_3}{(2\pi)^6}
\left(|{\cal A}_1|^2 + |{\cal A}_2|^2\right) \nonumber \\
&=& \frac{|g_\nu|^2}{4\mt^2 (2\pi)^3} \int d^2k_s
\frac{d^3k_2}{w_2} \frac{d^3k_3}{w_3} \delta^4(k_1+k_s-k_2-k_3)
\nonumber \\ && \hspace{1in} \times
\left(\frac{k_s^2}{\mB^2}(c_1^2+c_3^2)w_2 + 
2 c_1 c_2 \frac{\bk_s \cdot \bk_2}{\mB} + c_2^2 w_2\right) e^{-k_s^2/\mt^2} \ .
\eea
The $d^3k_2$ and $d^3k_3$ integrals can be evaluated using
\bea
\int \frac{d^3k_2}{w_2} \frac{d^3k_3}{w_3} k_2^\mu \delta^4(k-k_2-k_3)
 = k^\mu \pi \left(1-\frac{m_\phi^2}{k^2}\right)^2 \Theta(k^2 - m_\phi^2) \ .
\eea
This can most easily be shown in the centre of outgoing momentum frame, and
then Lorentz transformed to a general frame. We expand the remaining
$d^2k_s$ integral as a power series in $m_\phi/\mB$.
\bea
\Gamma &=& \frac{|g_\nu|^2 \mB}{16\pi\mt^2} 
\int_0^{\sqrt{\mB^2-m_\phi^2}} k_s dk_s 
\left[c_2^2 + \left(c_1^2+c_3^2 + 2c_1c_2\right) \frac{k_s^2}{\mB^2}\right]
\left(1 -\frac{m_\phi^2}{\mB^2-k_s^2}\right)^2 e^{-k_s^2/\mt^2}
\nonumber \\
&=& \mB \frac{|g_\nu|^2}{32\pi}\left\{ c_2^2(1-e^{-\lambda})
- (c_1^2 + c_3^2 + 2c_1 c_2)[\lambda^{-1} + e^{-\lambda}(1-\lambda^{-1})]
\right\} + O\left(\frac{m_\phi^2}{\mB^2}\right) \ .
\eea
where $\lambda = (\mB/\mt)^2 \sim 1$. 

The massive bound state lifetime is then
\be
\tau \sim (|g_\nu|^2 \mB)^{-1} \ ,
\ee
which is small.

\section{Zero Mode Scattering on Nearby Strings} 
\label{ap vort}

In addition to plasma interactions, it is also possible for massless
fermion charge-carriers to decay by interacting with currents on other strings.
This will be most significant when the separation of the
strings is small.

We will consider two infinite
straight strings with opposite windings, running parallel to each
other. We define $2\bR$ to be the displacement of the second
string, with $\bR$ orthogonal to the z-axis. This calculation will also be
of relevance to string loops, $R$ is then the radius of
the corresponding loop. For a typical vorton (a loop stabilised by a current)
$R \sim 10 - 100 \ms^{-1}$.

Since the second string has opposite winding to those considered in
section~\ref{spectrum}, the solution \bref{chians0} is not
valid. Instead
\be
\loarrow{U}_{0\bk} \propto \frac{\mt}{\sqrt{\pi L}} \bmat{c} 0 \\ 1 \emat 
\exp\left(-\int \frac{M}{\ms} f + \frac{a}{2\rho_\sstR} d\rho_\sstR \right)
 e^{ik_z z}
\label{chiansR}
\ee
should be used, with $\rho_\sstR$ a radial coordinate centred on
$2\bR$ rather than the origin. These states have $k_z <0$, and
opposite chirality to \bref{chians0}.

The cross section for direct interactions between currents on
different strings will be proportional to the square of the overlap of
their wavefunctions, which is of order $\exp(-4 \mt R)$. 
Interactions involving the exchange of a virtual particle
do not suffer from this suppression, and provide the dominant decay
channel. We will consider the exchange of a virtual Higgs particle
$\roarrow{\nu}^c_{\! str} + \loarrow{\nu}^c_{\! str}
\rightarrow \nu + \bar\nu$, 
as shown in figure~\ref{Feynman fig}d (the arrows indicate
the direction of the bound state momenta along the strings).
\bea
{\cal A}&=& \bigl< \roarrow{\nu}^c_{\! str}(\bk_1)
\loarrow{\nu}^c_{\! str}(\bk_2) | S \ket{\nu(\bk_3) \, \bar\nu(\bk_4)}
\nonumber \\
&=& 2 |g_\nu|^2 \sum_\bk\int d^4x \int d^4y \, 
\left(\roarrow{U}_{0\bk_1}^\dagger(\bx) i\sig{2} U_{\bk_3}(\bx) e^{i(w_1-w_3)t}
\loarrow{U}_{0\bk_2}^\dagger(\bx') i\sig{2} V_{\bk_4}^*(\bx') e^{i(w_2-w_4)t'}
\right. \nonumber \\ &&\ \ \left.
{}-\roarrow{U}_{0\bk_1}^\dagger(\bx) i\sig{2} V_{\bk_4}^*(\bx') e^{i(w_1-w_2)t}
\loarrow{U}_{0\bk_2}^\dagger(\bx') i\sig{2} U_{\bk_3}(\bx) e^{i(w_2-w_3)t'} 
\right) G(x-x') \ .
\label{vor amp1}
\eea
Substituting \bref{chians0}, \bref{greens}, \bref{str states} and
\bref{chiansR}, with $\roarrow{U}_{0\bk} = U_{0\bk}$, and evaluating the
integrals gives
\bea
{\cal A}&=& \frac{4 \pi |g_\nu|^2 c_0^2}{LV\mt^2}
\left\{ \sqrt{1-\frac{k_{3z}}{w_3}} \sqrt{1+\frac{k_{4z}}{w_4}} 
e^{i\beta_3+i\beta_4} {\cal I}(k_{1z}-k_{3z},w_1-w_3) e^{2i\bk_4\cdot \bR}
\right. \nonumber \\ && \left.
{}- \sqrt{1+\frac{k_{3z}}{w_3}} \sqrt{1-\frac{k_{4z}}{w_4}}
{\cal I}^*(k_{2z}-k_{3z},w_2-w_3) e^{2i\bk_3\cdot \bR} \right\} 
\delta^{(t,z)} (k_1 + k_2 - k_3 - k_4) \hspace{.4in}
\label{vor amp2}
\eea
where
\be
{\cal I}(k_z,w) = \int \frac{d^2k}{m_\phi^2 + k_z^2 - w^2 + k_T^2}
\exp\left(-\frac{(\bk+\bk_3)_T^2}{2\mt^2} -\frac{(\bk-\bk_4)_T^2}{2\mt^2}
-2i\bk\cdot \bR \right) \ .
\ee
Applying the convolution theorem gives
\bea
{\cal I}(k_z,w) &=& 2\mt^2 \int d^2X
K_0\left(2\sqrt{m_\phi^2+ k_z^2 - w^2} |\bR-{\bf X}|\right)
 \nonumber \\ && \ \ \times 
e^{-X^2 \mt^2} e^{i(\bk_3-\bk_4)\cdot {\bf X}} 
e^{-(\bk_4+\bk_3)_T^2/4\mt^2} \ .
\label{vorint1}
\eea
The main contribution from the integrand occurs around
$X=0$. Since the modified Bessel function $K_0$ varies slowly in this
region, \bref{vorint1} can be approximated by putting $X=0$ in the
argument of the Bessel function. This gives
\be
{\cal I}(k_z,w) \approx 2\pi e^{-k_{3T}^2/2\mt^2} e^{-k_{4T}^2/2\mt^2}
K_0\left(2\sqrt{m_\phi^2+ k_z^2 - w^2} R\right) \ .
\label{vorint2}
\ee
The exponential factors come from non-conservation of transverse
momentum at each string and the Bessel function is a result of the
limited range of the virtual Higgs particle.

Substituting \bref{vorint2} into \bref{vor amp2}, squaring and multiplying
by $L/{\cal T} v_{\rm rel}$ gives the cross section. $v_{\rm rel}=2$ and so,
\bea
\sigma &=& \frac{2 |g_\nu|^4 c_0^4}{(2\pi)^4 \mt^4}
\int dw_3 d\Omega_3 dw_4 d\Omega_4 \delta^{(t,z)} (k_1 + k_2 - k_3 - k_4)
w_3 w_4 e^{-k_{3T}^2/\mt^2} e^{-k_{4T}^2/\mt^2}
\nonumber \\ &&
\left\{ (w_3-k_{3z})(w_4+k_{4z}) 
K_0^2\left(2\sqrt{m_\phi^2+ (k_{1z} - k_{3z})^2 - (w_1 - w_3)^2} R\right) 
+ (k_3 \longleftrightarrow k_4)
\right. \nonumber \\ && 
- 2k_{3T}k_{4T} \mbox{Re}\left( e^{i(\beta_3 - 2\bk_3 \cdot \bR)}
e^{i(\beta_4 + 2\bk_4 \cdot \bR)}\right) 
K_0\left(2\sqrt{m_\phi^2 + (k_{1z} - k_{3z})^2 - (w_1 - w_3)^2} R\right) 
\nonumber \\ && \left.
K_0\left(2\sqrt{m_\phi^2 + (k_{1z} - k_{4z})^2 - (w_1 - w_4)^2} R\right)
\right\} \ .
\label{vor sig1}
\eea 

The cross section gives the transition rate for unit incident
flux. The incident flux in this case is $L/v_{\rm rel}$, instead of
$V/v_{\rm rel}$ as in \bref{zm sig4}, because the incoming particles
are confined to the one-dimensional strings. This gives a
dimensionless cross section instead of the usual
$(\mbox{length})^{-2}$.

The Bessel function means that the first term of \bref{vor sig1} is
exponentially suppressed except when 
$m_\phi^2 + (k_{1z} - k_{3z})^2 - (w_1 - w_3)^2$ is small. We will use
the approximation $K_0(2z) = -\log(z)$ for $z<1$ and $K_0(2z) = 0$
elsewhere. In this region the transverse outgoing momenta are small,
and the gaussian factors can be dropped to first order. If 
$m_\phi R \ll 1$, the Higgs mass can also be neglected.

The integral of the second term of \bref{vor sig1} is equal to the integral
of the first. The third is of order $k_{3T}^2 k_{4T}^2$, which is
small in the region where the $K_0$ factors are not exponentially
small. Thus the third term of \bref{vor sig1} can be neglected. Using
the above approximations gives
\be
\sigma \sim \frac{|g_\nu|^4}{(\mt R)^4} \min(1, (w_{\rm CoM} R)^4) \ ,
\label{vor sig2}
\ee
where $w_{\rm CoM}$ is the incoming centre of mass energy.

If $m_\phi R$ is large the above approximations do not apply. This
occurs if $\Phi$ is a GUT boson, or the electroweak Higgs field at
high temperatures. In this case an asymptotic expansion of the $K_0$
factors can be used to estimate \bref{vor sig1}, and
\be
\sigma\sim {|g_\nu|^4 m_\phi \over \mt^4 R^3} e^{-4 m_\phi R} \ .
\ee

Taken at face value, \bref{vor sig2} gives a short lifetime for charge
carriers on a vorton. However, there are complications in directly
applying this result derived for straight strings to curved vortons.
The above calculation does not take into account conservation of this
angular momentum, as the system considered has no rotational
symmetry. For a perfectly circular loop we would expect angular
momentum to be conserved. Combined with energy conservation, this
would prevent massless modes on the string scattering into massive
modes. However, the fermionic spectrum has not been calculated for a
circular loop and there is no reason to expect the zero modes to
remain massless. Angular momentum  conservation should only be
considered in a consistent rotationally invariant calculation. The
calculation above works consistently with straight strings, thus
while \bref{vor sig2} may not be directly applicable to vorton
decay, it is relevant for interactions on non-circular loops and
scattering of currents on a string network.

\end{document}